\documentclass[aps,prx,twocolumn,superscriptaddress,nofootinbib]{revtex4-2}
\usepackage{amsmath,amssymb,graphicx}
\usepackage[colorlinks=true,linkcolor=blue,citecolor=blue,urlcolor=blue]{hyperref}

\begin{document}

\title{Comment on ``Provably Trainable Rotationally Equivariant Quantum Machine Learning''}

\author{Zhiming Xiao}
\author{Ting Li}
\affiliation{School of Communication and Information Engineering, Nanjing University of Posts and Telecommunications}

\date{\today}

\begin{abstract}
We comment on the article by West \textit{et al.}, ``Provably Trainable Rotationally Equivariant Quantum Machine Learning'' [PRX Quantum \textbf{5}, 030320 (2024)]. While the general framework is insightful, we identify a key inconsistency in the construction of the dynamical Lie algebra (DLA). Specifically, the fixed controlled-Z (CZ) gates applied to all nearest-neighbor qubits are treated as if they were parameterized gates, with generators expressed in terms of combinations of Pauli operators. We discuss the implications of this inclusion and encourage the authors to revisit their analysis using a corrected DLA formulation.
\end{abstract}
\maketitle

We commend the authors of Ref.~\cite{west2024provably} for their thoughtful and technically sophisticated contribution to symmetry-informed quantum machine learning . Their strategy, which leverages rotational symmetry and dynamical Lie algebra (DLA) analysis to address the barren plateau problem, is a promising step toward scalable, trainable quantum models.

However, we would like to highlight a critical issue in their DLA construction. In Appendix A of Ref.~\cite{west2024provably}, the authors include unparameterized gates---specifically, controlled-Z (CZ) gates---as generators in the DLA:
\begin{quote}
    ``The generators $H_j$ of the circuit are the operators that when exponentiated produce arbitrary rotations on the first $n_{\text{rad}}$ qubits, and CZ gates between nearest-neighbour qubits.''
\end{quote}
This formulation is inconsistent with the standard definition of the DLA as found in Refs.~\cite{ragone2023unified,fontana2023adjoint}, where the DLA is defined as the Lie closure of the algebra generated by parameterized gates only. But In their analysis, only parameterized gates are explicitly defined. Therefore, when the circuit includes unparameterized(fixed) gates, we argue that it is inappropriate to simply represent them as Hamiltonian evolutions and treat their generators as elements of the Lie closure. Instead, their effect on the parameterized gates should be carefully considered.

\begin{figure}[htbp]
    \centering
    \includegraphics[width=0.4\textwidth]{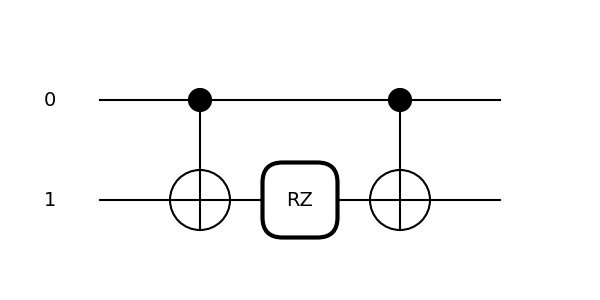} 
    \caption{Decomposition of ZZ-rotation  gates}
    \label{fig：Fig.1}
\end{figure}

We have an example to demonstrate this point.As depicted in Fig. \ref{fig：Fig.1}, the ZZ-rotation operator admits a decomposition utilizing CNOT and Z-rotation operators. For the purpose of computing the DLA, the generator corresponding to the CNOT gate is included as a basis element. Explicitly, we have
\[ CNOT = \exp\left[ i \frac{\pi}{4} (I_1 - Z_1) \otimes (I_2 - X_2) \right] \] from which we can get CNOT generator in terms of Pauli operators,\[cnot = (I_1 - Z_1) \otimes (I_2 - X_2)\]The Lie algebra associated with the circuit is then found by systematically computing the commutators of the basis elements until the algebra closes.And finally we found the DLA is composed of five elements:\begin{gather*}
    (I_1 - Z_1) \otimes (I_2 - X_2), 
    Z_2, 
    Y_2 - Z_1 \otimes Y_2, 
    -Z_2 + Z_1 \otimes Z_2, \\
    -X_2 + Z_1 \otimes X_2
\end{gather*}
This highlights a potential pitfall: initializing the DLA calculation with the generators of the constituent unparameterized gates (like CNOT from the decomposition) results in a DLA that does not accurately represent the dynamics of the overall parameterized circuit.

We suggest that the authors revisit their analysis using a corrected generator set consistent with standard Lie-algebraic frameworks. Such a revision will strengthen the theoretical guarantees claimed and maintain alignment with the broader literature on expressibility and barren plateaus.

One potential strategy involves analyzing how the collective action of CZ gates applied to all qubits transforms the Dynamic Lie Algebra (DLA) initially generated by the parameterized gates acting on the relevant 'radical' qubits of ref.~\cite{west2024provably}.This interaction is analogous to how a CNOT gate impacts the generator (e.g., ) of a Z-rotation gate.

We offer this comment in the spirit of constructive engagement and look forward to clarification from the authors.

\end{document}